\documentclass[twocolumn,trackchanges]{aastex62}
\shorttitle{The mass distribution of double neutron stars}
\shortauthors{Farrow, Zhu \& Thrane}

\usepackage{amsmath,units}
\usepackage{multirow}
\usepackage{subfigure}

\begin{document}

\title{The mass distribution of Galactic double neutron stars}

\correspondingauthor{Xing-Jiang Zhu}
\email{xingjiang.zhu@monash.edu}

\author{Nicholas Farrow}
\affiliation{School of Physics and Astronomy, Monash University, Clayton, VIC 3800, Australia}

\author[0000-0001-7049-6468]{Xing-Jiang Zhu}
\affiliation{School of Physics and Astronomy, Monash University, Clayton, VIC 3800, Australia}
\affiliation{OzGrav: Australian Research Council Centre of Excellence for Gravitational Wave Discovery, Clayton, VIC 3800, Australia}

\author{Eric Thrane}
\affiliation{School of Physics and Astronomy, Monash University, Clayton, VIC 3800, Australia}
\affiliation{OzGrav: Australian Research Council Centre of Excellence for Gravitational Wave Discovery, Clayton, VIC 3800, Australia}

\begin{abstract}
The conventional wisdom, dating back to 2012, is that the mass distribution of Galactic double neutron stars (DNSs) is well-fit by a Gaussian distribution with a mean of $1.33 M_\odot$ and a width of $0.09 M_\odot$.
With the recent discovery of new Galactic DNSs and GW170817, the first neutron star merger event to be observed with gravitational waves, it is timely to revisit this model.
In order to constrain the mass distribution of DNSs, we perform Bayesian inference using a sample of 17 Galactic DNSs effectively doubling the sample used in previous studies.
We expand the space of models so that the recycled neutron star need not be drawn from the same distribution as the nonrecycled companion. Moreover, we consider different functional forms including uniform, single-Gaussian, and two-Gaussian distributions.
While there is insufficient data to draw firm conclusions, we find positive support (a Bayes factor (BF) of 9) for the hypothesis that recycled and nonrecycled neutron stars have distinct mass distributions.
The most probable model --- preferred with a BF of 29 over the conventional model --- is one in which the recycled neutron star mass is distributed according to a two-Gaussian distribution, and the nonrecycled neutron star mass is distributed uniformly.
We show that precise component mass measurements of $\approx 20$ DNSs are required in order to determine with high confidence (a BF of 150) whether recycled and nonrecycled neutron stars come from a common distribution.
Approximately $60$ DNSs are needed in order to establish the detailed shape of the distributions.
\end{abstract}


\section{Introduction}\label{sec:intro}
The breakthrough discovery of GW170817 was the first ever detection of gravitational waves arising from a NS merger, $\unit[44]{Mpc}$ from Earth in the galaxy NGC4993 \citep{GW170817,170817optical}.
More binary neutron star events are expected when Advanced Laser Interferometer Gravitational-Wave Observatory \citep[LIGO;][]{aLIGO} and Advanced Virgo \citep{aVirgo} detectors resume observing with improved sensitivity this year \citep{LVCprospects}.
In our own Galaxy, the first DNS system, PSR B1913+16, was discovered in 1974 \citep{HulseTaylor75}.
The number of known Galactic DNS systems has increased to 18 so far \citep[see][and references therein]{BNSspin}, with three new discoveries published in 2018 \citep{Cameron17BNS,Lynch0509,BNS1946p20}.

The mass distribution of Galactic DNSs has been studied by numerous groups.
\citet{Ozel12} and \citet{Kiziltan13} fit the mass measurements for nine binaries to a Gaussian distribution.
This Gaussian model was used by \citet{TaylorCosmology12} to investigate the prospect of constraining cosmological parameters using gravitational-wave observations alone,
and by \citet{ZhuCBC13} to compute the gravitational-wave background from a population of NS mergers.
\citet{ThompsonSN12} demonstrated that the observed mass distribution contains information about the supernova explosion mechanism.
\citet{OzelARAA16} updated the initial Gaussian fit using measurements of two additional systems and found the most likely values of the mean and width to be $\mu=1.33 M_{\odot}$ and $\sigma=0.09 M_{\odot}$.

Since the discovery of GW170817, the mass distribution of Galactic DNSs has seen applications in a wider range of research.
Specifically, the conventional model of \citet{OzelARAA16} was used (1) as an astrophysical prior to place constraints on the NS equation of state using data from GW170817 \citep{DeBrown18},
(2) to argue that GW170817 is unlikely to have come from a population of DNSs like those found in our Galaxy \citep{Pankow18},
(3) to investigate the possibility of detecting a population of postmerger signals from NS mergers \citep{YangBNSmerger},
and (4) as a prior to argue that fast radio bursts may be associated with NS mergers resulting in prompt collapse \citep{FRBbns18}.
Additionally, \citet{MaBNSremnant} adopted the single-Gaussian models derived in \citet{Ozel12} and \citet{Kiziltan13} to investigate the prospect of forming supramassive NSs from NS mergers.
\citet{BNSspin} developed a framework to infer DNS population properties through gravitational-wave observations, assuming the masses and spin periods of Galactic DNSs are representative of the merging population.

In light of the new discoveries of Galactic DNSs from radio astronomy and the ever-expanding catalog of gravitational-wave events, we revisit the problem of inferring the mass distribution of Galactic DNS systems. Apart from using a sample of observations that is nearly twice the size of earlier datasets, we expand on previous studies in several ways.
First, we enlarge the space of models so that the recycled NS and its nonrecycled companion do not necessarily follow the same mass distribution.
Second, we consider different functional forms for the mass distribution to include uniform, Gaussian, and two-Gaussian distributions.
Third, we account for the correlation between mass measurements for pulsars and their companions, which appears to be ignored in previous work.

This paper is organized as follows. In Section \ref{sec:data}, we review the mass measurements of Galactic DNS systems. In Section \ref{sec:method}, we describe the formalism for Bayesian inference and model selection. In Section \ref{sec:results}, we present and discuss our analysis results. Finally, we conclude in Section \ref{sec:conclu}.

\section{Masses of Galactic DNS}
\label{sec:data}
Among 18 Galactic DNS systems, 17 have reported measurements of the binary total mass, and 12 of which have masses measured for both component stars; see Table III in \citet{BNSspin} and references therein for details.
In this work, we divide the NSs in these binaries into two categories: recycled NSs and nonrecycled (`\textit{slow}') NSs, for which the masses are labeled as $m_r$ and $m_s$, respectively.

In the standard isolated binary formation channel \citep[e.g.,][]{Tauris17}, a recycled NS is the first born object that gets spun up to $\sim 10-100$ ms through an accretion/recycling process, whereas the second born NS quickly ($\sim$ Myr) spins down to a period of $\mathcal{O}(1)$ second after its birth.
The double pulsar PSR J0737$-$3039A/B is an excellent example, with pulsars A and B representing recycled and slow NSs respectively \citep{Burgay03,Lyne04}. All pulsars in the remaining 16 binaries are recycled NSs except PSR J1906+0746, which is a young slow pulsar \citep{vanLeeuwen15}.
Therefore, the companion of PSR J1906+0746 is included in the category of recycled NSs. Although the possibility of a massive white dwarf companion cannot be ruled out from radio observations, we take PSR J1906+0746 to be in a DNS system, as its orbital characteristics and mass estimates are consistent with other Galactic DNSs, especially the double pulsar \citep{J1906dpsr}.

\begin{figure*}
\centering
\begin{subfigure}
\centering\includegraphics[width=0.46\textwidth]{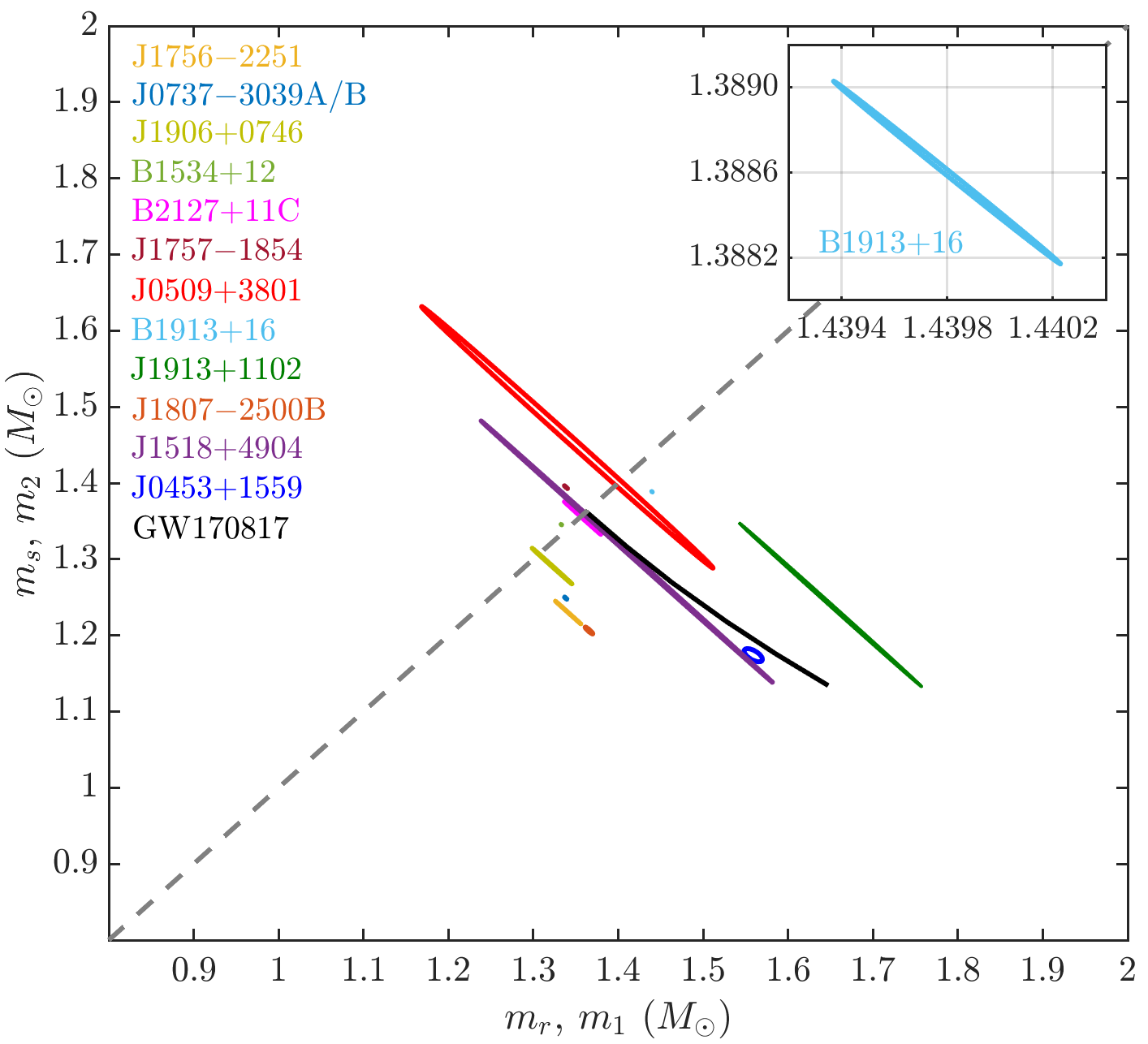}
\end{subfigure}%
\begin{subfigure}
\centering\includegraphics[width=0.46\textwidth]{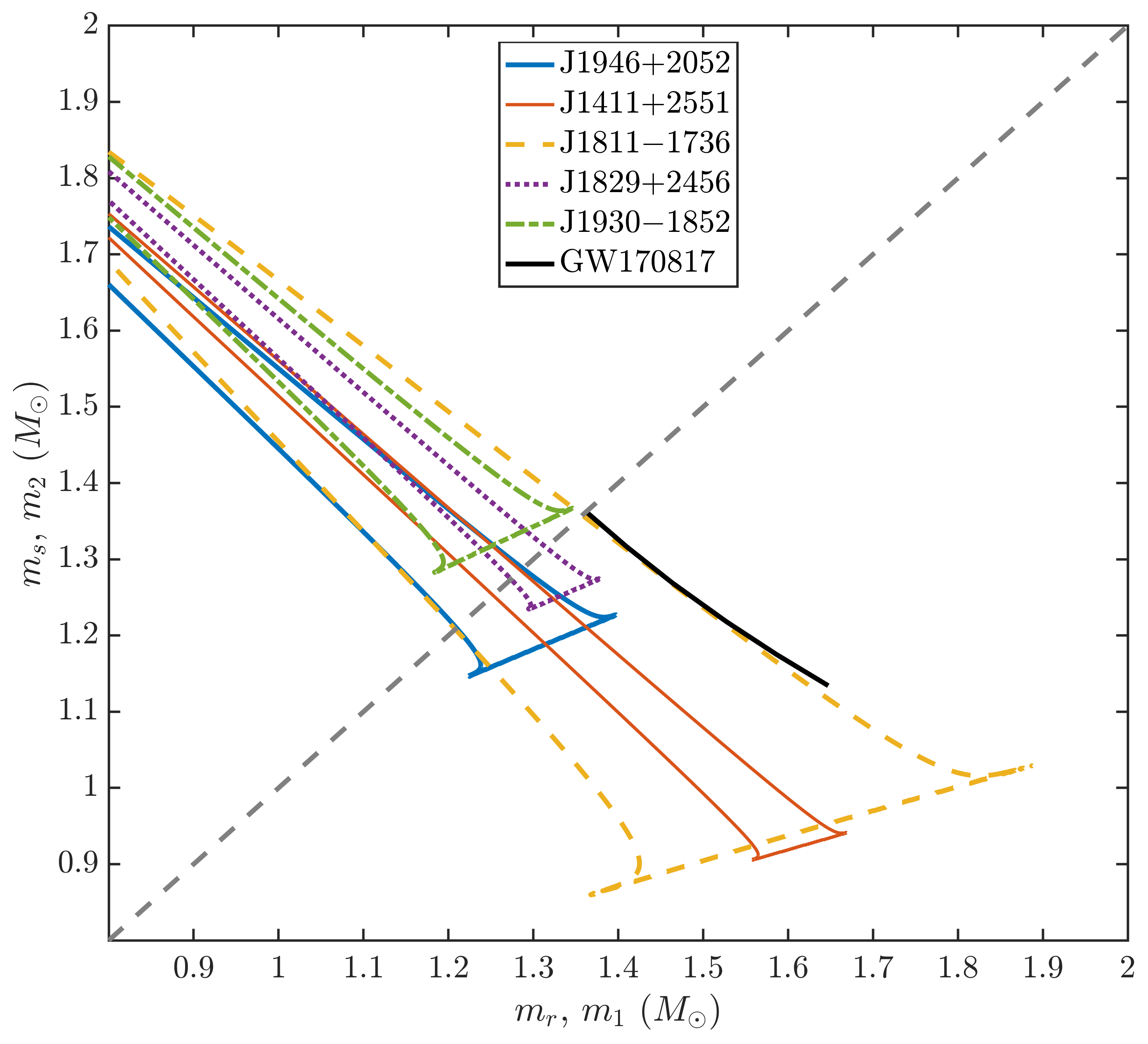}
\end{subfigure}
\caption{90\% confidence credible regions of recycled ($m_r$) and slow ($m_s$) NS masses for Galactic DNS systems: left---12 systems with measurements of both masses; the inset shows a 1000$\times$ zoomed-in look at the joint mass distribution for PSR B1913+16; right---5 binaries with only total mass measurements and an additional constraint that $|\sin i|\leq 1$ with $i$ being the orbital inclination angle. In both plots, we also show component mass ($m_{1}\geq m_{2}$) measurements of GW170817 \citep{170817property} for illustration purposes only (i.e., not used in the actual analysis).}\label{fig:m12BNS}
\end{figure*}

Note that PSR B2127+11C and PSR J1807$-$2500B are located in globular clusters, whereas the rest are in the Galactic disk.
It has been hypothesized that these two DNSs were formed from a different evolutionary scenario than the one described above~ \citep{B2127M15,Lynch12}.
In addition, the companion of PSR J1807$-$2500B could also be a massive white dwarf. Nevertheless, we include them in the analysis. As can be seen in Figure \ref{fig:m12BNS}, their masses are consistent with other DNS systems. Therefore, we do not expect such a choice to significantly bias our results.

Table \ref{tab:BNSmass} lists the measurements or constraints of the binary total mass $M_T$, mass of the recycled ($m_r$), and slow ($m_s$) NS for 17 binaries. To aid the comparison with gravitational-wave measurements, we also list values for the binary chirp mass, defined as
\begin{equation}
\mathcal{M}_c=\frac{(m_{1} m_{2})^{3/5}}{(m_{1}+m_{2})^{1/5}}\, ,
\end{equation}
and mass ratio $q=m_{2}/m_{1}\leq 1$.

\begin{table*}
\begin{center}
 \begin{tabular}{lcccccccc}
 \hline
  \hline
  Pulsar Name & $M_{T}$ ($M_{\odot}$) & $m_r$ ($M_{\odot}$) & $m_s$ ($M_{\odot}$) & $\mathcal{M}_{c}$ ($M_{\odot}$) & $q$ & $P_b$ (day) & $T_c$ (Gyr) & References \\
  \hline
  \multicolumn{9}{| c |}{Systems will merge within a Hubble time}\\
  \hline
  J1946+2052 & 2.50(4) & $< 1.35$ & $> 1.17$ & (1.05, 1.11) & (0.68, 1) & 0.078 & 0.046 & (1) \\
  J1756$-$2251 & 2.56999(6)  & 1.341(7) & 1.230(7) & 1.1178(3)  & 0.92(1) & 0.320 & 1.656 & (2) \\
J0737$-$3039A/B & 2.58708(16)  & 1.3381(7) & 1.2489(7) & 1.1253(1)  & 0.933(1) & 0.102 & 0.086 & (3) \\
 J1906+0746 & 2.6134(3) & 1.322(11) & 1.291(11)  & 1.1372(2) & (0.956, 1) & 0.166 & 0.308 & (4) \\
B1534+12  & 2.678463(4)  & 1.3330(2) & 1.3455(2)& 1.165870(2) & 0.9907(3) & 0.421 & 2.734 & (5) \\
B2127+11C  & 2.71279(13) & 1.358(10) & 1.354(10) &  1.18043(8)  & (0.975, 1) & 0.335 & 0.217 & (6) \\
 J1757$-$1854 & 2.73295(9) & 1.3384(9) & 1.3946(9) & 1.18930(4)  & 0.960(1) & 0.184 & 0.076 & (7) \\
 J0509+3801 & 2.805(3) & 1.34(8) & 1.46(8) & 1.215(5) &  (0.793, 1) & 0.380 & 0.574 & (8) \\
B1913+16 & 2.828378(7)  & 1.4398(2) & 1.3886(2) & 1.230891(5)  & 0.9644(3) & 0.323 & 0.301 & (9) \\
  J1913+1102  & 2.886(1) & 1.65(5) & 1.24(5) & 1.242(8)  & 0.75(5) & 0.206 & 0.473 & (10) \\
  \hline
  \multicolumn{9}{| c |}{Systems will not merge within a Hubble time}\\
     \hline
 J1807$-$2500B & 2.57190(73)  & 1.3655(21) & 1.2064(21) & 1.1169(3) & 0.883(3) & 9.957 & 1044 & (11) \\
 J1518+4904  & 2.7183(7) & 1.41(8) & 1.31(8) &  1.181(5) & (0.794, 1) & 8.634 & 8832 & (12) \\
  J0453+1559  & 2.733(4) & 1.559(5) & 1.174(4) & 1.175(2) & 0.753(5) & 4.072 & 1453 & (13) \\
  J1411+2551 & 2.538(22) & $<1.64$ & $>0.92$ & (1.05, 1.11) & (0.57, 0.95) & 2.616 & 466  & (14) \\
 J1811$-$1736  & 2.57(10) & $<1.75$ & $>0.91$ & (1.02, 1.17) & (0.58, 0.95) & 18.78 & 1794  & (15) \\
J1829+2456  & 2.59(2) & $<1.36$ & $>1.25$ & (1.08, 1.14) & (0.65, 1) & 1.176 & 55 & (16) \\
 J1930$-$1852 & 2.59(4) & $<1.32$ & $>1.30$ & (1.07, 1.15) & (0.58, 0.96) & 45.06 & $\sim 10^{5}$ & (17) \\
  \hline
  \hline
 \end{tabular}
 \end{center}
 \caption{Mass measurements of Galactic DNS systems: the binary total mass ($M_T$), the masses of the recycled NS ($m_r$) and the slow NS ($m_s$), binary chirp mass ($\mathcal{M}_c$), mass ratio ($q$), binary orbital period ($P_b$) and coalescence time $T_c$. Figures in parentheses are 1-$\sigma$ uncertainties in the last quoted digit of the mass measurement. For five systems without component mass measurements, we list the 99\% confidence upper and lower limits for $m_r$ and $m_s$, derived from their measured $M_T$ and mass functions: 0.268184 (J1946+2052), 0.1223898 (J1411+2551), 0.128121 (J1811$-$1736), 0.29413 (J1829+2456), 0.34690765 (J1930$-$1852). For $\mathcal{M}_{c}$ and for $q$, if the distribution is non-Gaussian or extended, 90\% confidence intervals are given. References: (1). \citet{BNS1946p20}; (2). \citet{Ferdman14}; (3). \citet{Kramer06Sci}; (4). \citet{vanLeeuwen15}; (5). \citet{Fonseca14}; (6). \citet{Jacoby06}; (7). \citet{Cameron17BNS}; (8). \citet{Lynch0509}; (9). \citet{Weisberg10}; (10). \citet{Ferdman17}; (11). \citet{Lynch12}; (12). \citet{Janssen08}; (13). \citet{Martinez15}; (14). \citet{BNS1411}; (15). \citet{Corongiu07}; (16). \citet{Champion05}; (17). \citet{Swiggum15}. In addition, $m_r$ and $m_s$ values of PSR J1518+4904 were taken from \citet{Tauris17} who cites a private communication with G. Janssen.}
  \label{tab:BNSmass}
\end{table*}

\begin{figure*}[!htb]
  \centering
	\includegraphics[width=\textwidth]{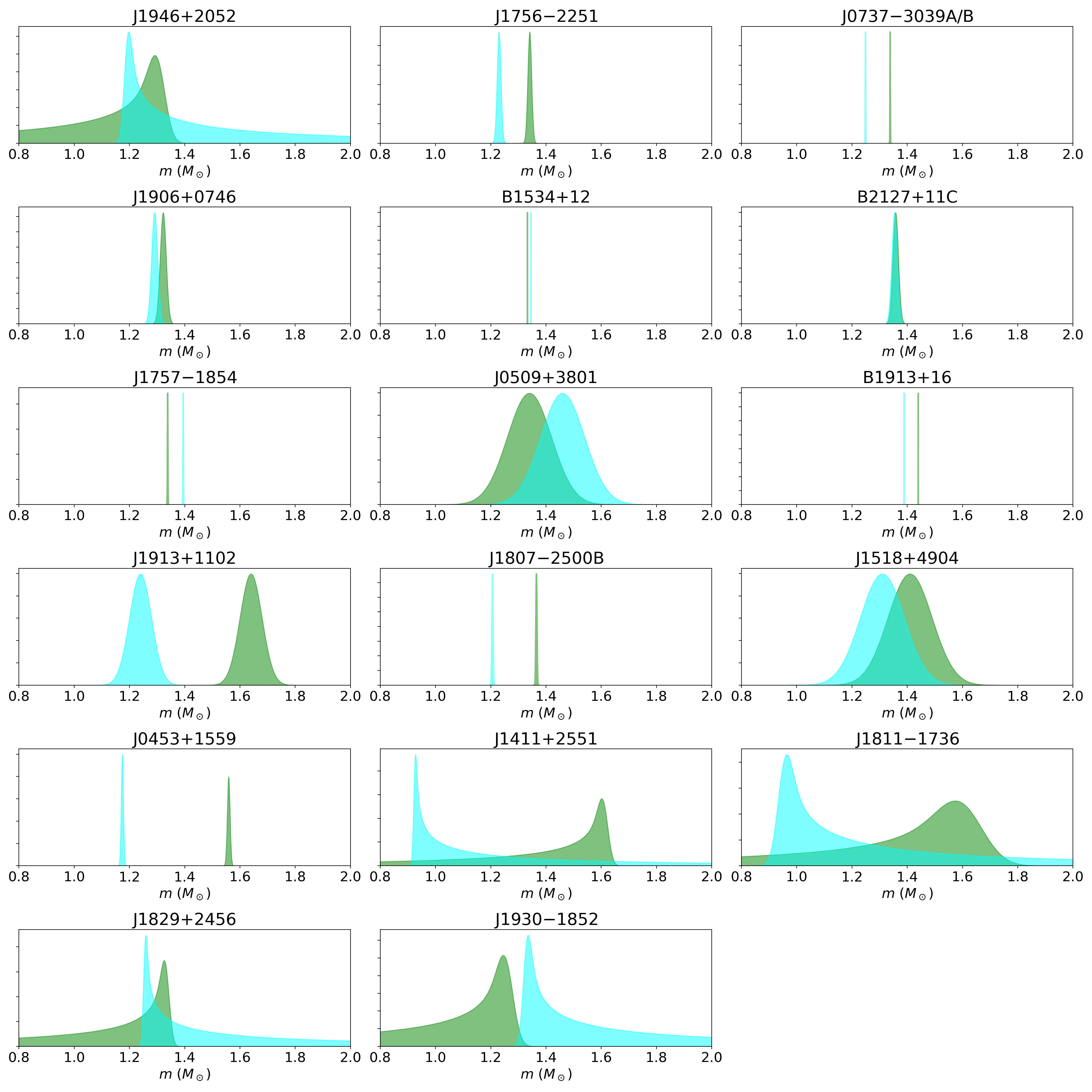}
  \caption{Marginalized mass distributions for recycled ($m_r$, green) and slow ($m_s$, cyan) NSs for 17 Galactic DNS systems.}
  \label{fig:pcSamples}
\end{figure*}

Figure \ref{fig:m12BNS} shows the joint distribution of $(m_{r},m_{s})$ for 17 Galactic DNS systems, along with the mass constraints of GW170817 \citep{170817property}. Figure \ref{fig:pcSamples} shows the marginalized distributions of $m_r$ and $m_s$. We briefly describe these mass distributions below; details of NS mass measurements through pulsar timing can be found in \citet{Stairs03}.

As can be seen in Figure \ref{fig:m12BNS}, the measurements of $m_r$ and $m_s$ are anticorrelated due to the constraint on $M_T$.
In the left plot are 12 systems for which precise component mass measurements are available, whereas on the right are another 5 binaries with only total mass measurements.
For the 12 systems, $(m_{r},m_{s})$ follows a bivariate normal distribution\footnote{This is a reasonable approximation to the original posterior distribution derived from radio pulsar timing observations; see, e.g., Figure 2 in \citet{Lynch0509}.} with a covariance of $(\sigma_{T}^2-\sigma_{r}^2-\sigma_{s}^2)/2$, where $\sigma_{T}$, $\sigma_{r}$ and $\sigma_{s}$ are the measurement errors quoted in parentheses in Table \ref{tab:BNSmass} for $M_{T}$, $m_{r}$ and $m_{s}$, respectively.

For the five systems shown in the right panel of Figure \ref{fig:m12BNS}, the joint distribution of $(m_{r},m_{s})$ is determined by measurements of $M_{T}$ and the mass function
\begin{equation}\label{eq:pdfm12mt}
\left.\begin{aligned}
p(m_{r},m_{s}) &\propto \exp\left[-\frac{(m_{r}+m_{s}-\hat{M}_{T})^2}{2\sigma_{T}^2}\right] \\
& \times \int_{-1}^{1} \delta (f-\hat{f})\text{d}(\cos i)\\ & \hspace{-2cm} \propto \frac{\exp\left[-\frac{(m_{r}+m_{s}-\hat{M}_{T})^2}{2\sigma_{T}^2}\right]}{\frac{3m_{s}^3}{(m_{r}+m_{s})^2}\left[\frac{\hat{f}(m_{r}+m_{s})^{2}}{m_{s}^3}\right]^{\frac{1}{3}}\left[1-\frac{\hat{f}^{2/3}(m_{r}+m_{s})^{4/3}}{m_{s}^2}\right]^{\frac{1}{2}}}\, ,
\end{aligned}\right.
\end{equation}
where $\hat{f}$ and $\hat{M}_{T}$ are measured mean values of the mass function and total mass, respectively. The mass function is defined as
\begin{equation}
f=\frac{(m_{c} \sin i)^3}{M_{T}^2}=\left(\frac{2\pi}{P_b}\right)^2\frac{(a\sin i)^3}{G}\, ,
\label{eq:massfunc}
\end{equation}
where $P_b$, $a$ and $i$ are the orbital period, semimajor axis, and orbital inclination angle, respectively; $m_c$ is the mass of the pulsar's companion, which is replaced with $m_s$ in Equation (\ref{eq:pdfm12mt}) because for the systems in question, the companion is a slow NS. As in \citet{Alsing18}, we ignore the measurement uncertainty of mass function to derive the last line in Equation (\ref{eq:pdfm12mt}).

\section{Bayesian Inference}
\label{sec:method}
\subsection{Constraining the Shape of NS Mass Distributions with Hyperparameters}
Our starting point is the measurement of NS masses from radio or gravitational-wave observations\footnote{So far there is only one DNS system (GW170817) that is identified via gravitational waves. Its mass constraints are presented for $m_1$ and $m_2$. Since we are interested in the distribution of $m_r$ and $m_s$, we focus on radio pulsar observations in this work.}.
These previously published constraints on NS masses are {\em posterior} distributions, which we denote as
\begin{align}
  p(m_r^i, m_s^i | d^i) ,
\end{align}
where $i$ refers to the $i$th binary.
Formally, this is the joint posterior on the recycled NS mass $m_r$ and the slow NS mass $m_s$ given some data $d$, as illustrated in Figure \ref{fig:m12BNS}.

We use the posterior distributions from published NS mass constraints to study the population properties of NS using hierarchical Bayesian inference.
We introduce priors for $m_r$ and $m_s$, which are conditional on hyperparameters $(\Lambda_r, \Lambda_s) \in \Lambda$
\begin{align}
& \pi(m_r^i | \Lambda_r, H) \\
& \pi(m_s^i | \Lambda_s, H) .
\end{align}
The hyperparameters $(\Lambda_r, \Lambda_s)$ describe the shape of the $(m_r,m_s)$ priors given a hypothesis $H$.
For example, later we consider a hypothesis in which $\pi(m_r | \Lambda_r, H)$ is a Gaussian distribution in $m_r$.
In this case, $\Lambda_r$ consists of  hyperparameters for the mean $\mu_r\in\Lambda_r$ and width $\sigma_r\in\Lambda_r$ of the distribution.

Our first goal is to calculate (hyper)posterior distributions for $(\Lambda_r,\Lambda_s)$.
Given a data set with $N$ binary mass measurements, the hyperposterior distribution is given by~\citep{ThraneTalbot}
\begin{align}\label{eq:p_eht}
p(\Lambda|\{d\}) = & 
\frac{1}{{\cal Z}_\Lambda}
\prod_i^N
\int dm_r^i \int dm_s^i \,
{\cal L}(d^i|m_r^i, m_s^i) \nonumber\\
& \pi(m_r^i|\Lambda_r) 
\pi(m_s^i|\Lambda_s) \pi(\Lambda_r)
\pi(\Lambda_s) .
\end{align}
Here $\pi(\Lambda_r)$ and $\pi(\Lambda_s)$ are the priors of our hyperparameters, which we take to be flat.
The term ${\cal Z}_\Lambda$ is the (hyper)evidence
\begin{align}
\label{eq:ZLambda}
{\cal Z}_\Lambda \equiv & 
\int d\Lambda_s \int d\Lambda_r \,
\prod_i^N
\int dm_r^i \int dm_s^i \nonumber\\
& {\cal L}(d^i|m_r^i, m_s^i)
\pi(m_r^i|\Lambda_r) 
\pi(m_s^i|\Lambda_s) \pi(\Lambda_r)
\pi(\Lambda_s)\, .
\end{align}
The variable ${\cal L}(d^i|m_r^i, m_s^i)$ is the likelihood function of the data given $(m_r^i,m_s^i)$.
We do not have direct access to this likelihood, but it is related to the posterior distribution
\begin{align}\label{eq:L_eht}
{\cal L}(d^i|m_r^i,m_s^i) = \frac{{\cal Z}_0^i}{\pi_0(m_r^i, m_s^i)} p(m_r^i, m_s^i|d^i) ,
\end{align}
where $\pi_0(m_r,m_s)$ is the initial prior used to derive the mass posterior (assumed here to be flat).
Meanwhile, 
\begin{align}
{\cal Z}_0 \equiv \int dm_r \int dm_s \,
{\cal L}(d|m_r, m_s) \pi_0(m_r,m_s) ,
\end{align}
is the initial evidence.
Without access to the raw radio data $d$, we do not know ${\cal Z}_0$, but this factor can be ignored since it will ultimately cancel, via either normalization of the hyperposterior or the construction of a Bayesian evidence ratio.
For the sake of readability, we therefore set ${\cal Z}_{0}^i=1$ henceforth.

Plugging Equation \ref{eq:L_eht} into \ref{eq:p_eht}, we obtain
\begin{align}\label{eq:Z_eht}
p(\Lambda|\{d\}) = & \frac{1}{{\cal Z}_\Lambda} 
\prod_i^N 
\int dm_r^i \int dm_s^i \,
p(m_r^i, m_s^i|d^i) \nonumber\\
& \frac{\pi(m_r^i|\Lambda_r)\pi(m_s^i|\Lambda_s)} 
{\pi_0(m_r^i,m_s^i)} \pi(\Lambda_r)
\pi(\Lambda_s)\, .
\end{align}
Using Eq.~\ref{eq:Z_eht}, we construct  posteriors on the hyperparameters that describe the shape of NS mass distributions.

In this work, we employ the \textsc{pymultinest} implementation \citep{PyMultiNest} of the \textsc{MultiNest} algorithm \citep{MultiNest} for the stochastic sampling of posteriors and the evidence calculation. Our codes are publicly available on GitHub, \url{https://github.com/NicholasFarrow/GalacticDNSMass}.

\subsubsection{Functional Forms of NS Mass Distributions}
We consider three functional forms for the distributions of $m_r$ and $m_s$. The first is a uniform distribution.
It comprises two parameters, the lower and upper bound: $\{m_{r}^{l}, m_{r}^{u}\} \in \Lambda_{r}$, and $\{m_{s}^{l}, m_{s}^{u}\} \in \Lambda_{s}$.

The second form is a Gaussian distribution, which is commonly used in previous studies on the NS mass distributions \citep[e.g.,][]{Ozel12,Kiziltan13}. In this case, the probability distribution is given by:
\begin{equation}
\pi(m|\Lambda) = \pi(m|\{\mu,\sigma\}) = \frac{1}{\sigma\sqrt{2\pi}}\exp\left[-\left(\frac{m-\mu}{\sqrt{2}\sigma}\right)^{2}\right]\, ,
\end{equation}
where $\mu$ and $\sigma$ correspond to the mean and width of the distribution, respectively.

A natural extension to the Gaussian form is that the distribution has two distinct peaks, leading to the two-Gaussian distribution:
\begin{eqnarray}
\pi(m|\Lambda)&=&\pi(m|\{\mu_1,\sigma_1,\mu_2,\sigma_2,\alpha\}) = \frac{\alpha}{\sigma_1\sqrt{2\pi}} \\ && \hspace{-12mm} \times \exp\left[-\left(\frac{m-\mu_1}{\sqrt{2}\sigma_1}\right)^{2}\right] + \frac{1-\alpha}{\sigma_2\sqrt{2\pi}}\exp\left[-\left(\frac{m-\mu_2}{\sqrt{2}\sigma_2}\right)^{2}\right]\, .  \nonumber 
\end{eqnarray}
Here $\alpha$ and $1-\alpha$ give the weight of the first (with mean $\mu_1$ and width $\sigma_1$) and second (with mean $\mu_2$ and width $\sigma_2$) peak, respectively.

We adopt uniform priors for all parameters. Table \ref{tab:PriorRanges} lists the prior ranges, which apply to both $m_r$ and $m_s$. The Gaussian and two-Gaussian distributions are truncated at 0.8 and 2 $M_{\odot}$ to reflect these prior ranges.
Note that the precisely measured NS masses are well within this range, so our results depend only weakly on the choice of this prior range.

\begin{table}[!htb]
\centering
 \begin{tabular}{lccc}
  \hline
 Form & Parameter & Min & Max \\
  \hline
  \multirow{2}{*}{Uniform} & $m^l$ & 0.8 & 2\\
   & $m^u$ & $m^l$ & 2\\
   \hline
   \multirow{2}{*}{Gaussian} & $\mu$ & 0.8 & 2 \\
   & $\sigma$ & 0.005 & 0.5 \\
   \hline
  \multirow{5}{*}{Two-Gaussian} & $\mu_{1}$ & 0.8 & 2 \\
   & $\mu_{2}$  & $\mu_1$ & 2  \\
   & $\sigma_{1}$ & 0.005 & 0.5  \\
   & $\sigma_{2}$ & 0.005 & 0.5  \\
   & $\alpha$ & 0 & 1 \\
  \hline
 \end{tabular}
 \caption{Prior ranges for three functional forms of NS mass distribution. Mass parameters are in unit of $M_{\sun}$.}
  \label{tab:PriorRanges}
\end{table}

\subsection{Model Selection}
The second goal of this work is to differentiate hypotheses of NS mass distributions.
Using Bayes' theorem, we can compute the ratio of posterior probability between the two hypotheses $H_1$ and $H_2$ as follows:
\begin{equation}
\mathcal{O}=\frac{P(H_1|\{d\})}{P(H_2|\{d\})} = \frac{\mathcal{Z}_{\Lambda}(H_1)P(H_1)}{\mathcal{Z}_{\Lambda}(H_2)P(H_2)}\, ,
\end{equation}
where $P(H_1)$ and $P(H_2)$ are the prior probability for hypotheses $H_1$ and $H_2$ respectively; $\mathcal{Z}_{\Lambda}(H_1)$ and $\mathcal{Z}_{\Lambda}(H_2)$ are the Bayesian evidence for hypotheses $H_1$ and $H_2$, respectively; see Equation (\ref{eq:ZLambda}). The quantity defined above is usually called the odds ratio.
Assuming equal prior probability, $P(H_1)=P(H_2)$, model selection is performed by computing the Bayes factor (BF)
\begin{equation}
\text{BF}^{1}_{2} = \frac{\mathcal{Z}_{\Lambda}(H_1)}{\mathcal{Z}_{\Lambda}(H_2)}
\end{equation}

We follow the BF interpretation outlined by \citet{Raftery95}: $0<\text{BF}^{1}_{2}<3$ indicates that the support for $H_1$ is `worth not more than a bare mention', $3<\text{BF}^{1}_{2}<20$ indicates positive support, $20<\text{BF}^{1}_{2}<150$ means the data strongly favor $H_1$, and $\text{BF}^{1}_{2}>150$ indicates very strong support.
We use BF for small BFs and its natural logarithm, ln(BF), for large values. We choose a threshold for BF of 150, corresponding to a ln(BF)=5, as required for confident model selection.

Here we wish to compare two hypotheses: $\mathcal{A}$ -- recycled and slow NSs follow an identical mass distribution, and $\mathcal{B}$ -- they are drawn from two distinct populations. For hypothesis $\mathcal{A}$, there are three possibilities: $m_r$ and $m_s$ follow 1) a uniform distribution, or 2) a Gaussian distribution, or 3) a two-Gaussian distribution. In contrast, there are nine possibilities for hypothesis $\mathcal{B}$, with $m_r$ and $m_s$ each following any one of the three distributions. The evidence for hypotheses $\mathcal{A}$ and $\mathcal{B}$ is
\begin{eqnarray}
&&\mathcal{Z^{A}}=\frac{1}{3}\left(\mathcal{Z}^{\mathcal{A}}_{\text{u}}+\mathcal{Z}^{\mathcal{A}}_{\text{g}}+\mathcal{Z}^{\mathcal{A}}_{\text{t}}\right)\, , \\ && \mathcal{Z^{B}}=\frac{1}{9}\left(\mathcal{Z}^{\mathcal{B}}_{\text{uu}}+\mathcal{Z}^{\mathcal{B}}_{\text{ug}}+\mathcal{Z}^{\mathcal{B}}_{\text{ut}}+\dots +\mathcal{Z}^{\mathcal{B}}_{\text{tt}}\right)
 \, ,
\end{eqnarray}
where u, g, and t in the subscripts denote uniform, Gaussian, and two-Gaussian distributions, respectively. Here the notation works as follows: $\mathcal{Z}^{\mathcal{A}}_{\text{u}}$ is the evidence for the case that $m_{r}$ and $m_{s}$ are drawn independently from an identical uniform distribution, whereas $\mathcal{Z}^{\mathcal{B}}_{\text{uu}}$ is the evidence for the case that $m_{r}$ and $m_{s}$ follow two uniform distributions with different parameters.

To quantitatively distinguish the support for each subhypothesis within the two hypotheses (three subhypotheses for $\mathcal{A}$, and nine subhypotheses for $\mathcal{B}$), we also compute the relative evidence (RE), defined as the ratio between the evidence of each subhypothesis and the total evidence of the hypothesis\footnote{This is equivalent to the relative model probability used in \citet{FarrBHmass11}, who performed model selection for the mass distribution of stellar-mass black holes.}. For example, the RE for subhypothesis u under the hypothesis $\mathcal{A}$ is
\begin{equation}
\text{RE}^{\mathcal{A}}_{\text{u}} = \frac{\mathcal{Z}^{\mathcal{A}}_{\text{u}}}{(\mathcal{Z}^{\mathcal{A}}_{\text{u}}+\mathcal{Z}^{\mathcal{A}}_{\text{g}}+\mathcal{Z}^{\mathcal{A}}_{\text{t}})} \, .
\end{equation}

\subsection{Posterior Predictive Distributions}
Using posterior distributions of hyperparameters $(\Lambda_r,\Lambda_s)$, we can derive a posterior predictive distribution (PPD)
\begin{align}
p_{\Lambda_r}(m_r) = & \int \text{d}\Lambda_r \,
p(\Lambda_r | \{d\}) \pi(m_r|\Lambda_r)\, ,
\\
p_{\Lambda_s}(m_s) = & \int \text{d}\Lambda_s \,
p(\Lambda_s | \{d\})
\pi(m_s|\Lambda_s)\, .
\end{align}

The PPD is our best guess for the updated prior on $(m_r,m_s)$ given the data $\{d\}$.
In some instances, it is useful to convert into a PPD for related variables such as $M_T=m_r+m_s$.
This is accomplished through convolution
\begin{align}
p_\Lambda(M_{T}) =  \int \text{d} m_r
p_{\Lambda_r}(m_r)p_{\Lambda_s}(M_{T}-m_r)\, .
\end{align}
The PPD for mass ratio $q\in (0, 1]$ can be obtained as follows
\begin{align}
p_\Lambda(q) = & \frac{1}{2} \left [\int \text{d} m_r
p_{\Lambda_s}(q m_r)p_{\Lambda_r}(m_r) m_{r} \right. \nonumber \\ & \left.
 + \int \text{d} m_s
p_{\Lambda_r}(q m_s)p_{\Lambda_s}(m_s) m_{s} \right]\, .
\end{align}

\section{Results}
\label{sec:results}
\begin{table}[!htb]
\centering
 \begin{tabular}{l|c|c|c|c}
  \hline
 Subhypothesis &  & $\mathcal{Z}^{\mathcal{A}}_{\text{u}}$ & $\mathcal{Z}^{\mathcal{A}}_{\text{g}}$ & $\mathcal{Z}^{\mathcal{A}}_{\text{t}}$ \\
  \hline
  & RE & 0.027 & 0.597 & 0.376 \\
  \hline
  $\mathcal{Z}^{\mathcal{B}}_{\text{uu}}$ & 0.197 & & & \\
   $\mathcal{Z}^{\mathcal{B}}_{\text{ug}}$ & 0.011 & & & \\
   $\mathcal{Z}^{\mathcal{B}}_{\text{ut}}$ & 0.002& & & \\
   $\mathcal{Z}^{\mathcal{B}}_{\text{gu}}$ & 0.087 & & & \\
   $\mathcal{Z}^{\mathcal{B}}_{\text{gg}}$ & 0.004 & & & \\
   $\mathcal{Z}^{\mathcal{B}}_{\text{gt}}$ & 0.001 & & & \\
   $\mathcal{Z}^{\mathcal{B}}_{\text{tu}}$ & 0.643 & & 28.6 & \\
   $\mathcal{Z}^{\mathcal{B}}_{\text{tg}}$ & 0.040 & & & \\
   $\mathcal{Z}^{\mathcal{B}}_{\text{tt}}$ & 0.015 & & & \\
  \hline
 \end{tabular}
 \caption{For \textit{all} Galactic DNS systems: relative evidences (RE) for Subhypotheses in hypotheses $\mathcal{A}$ and $\mathcal{B}$, and the BF between the Best Subhypothesis in $\mathcal{A}$, and the Best Subhypothesis in $\mathcal{B}$. On average, we find $\mathcal{Z^{B}}/\mathcal{Z^{A}}=8.9$.}
  \label{tab:evidenceBF}
\end{table}

\begin{table}[!htb]
\centering
 \begin{tabular}{l|c|c|c|c}
  \hline
 Subhypothesis &  & $\mathcal{Z}^{\mathcal{A}}_{\text{u}}$ & $\mathcal{Z}^{\mathcal{A}}_{\text{g}}$ & $\mathcal{Z}^{\mathcal{A}}_{\text{t}}$ \\
  \hline
  & RE & 0.087 & 0.486 & 0.427 \\
  \hline
  $\mathcal{Z}^{\mathcal{B}}_{\text{uu}}$ & 0.155 & & & \\
   $\mathcal{Z}^{\mathcal{B}}_{\text{ug}}$ & 0.008 & & & \\
   $\mathcal{Z}^{\mathcal{B}}_{\text{ut}}$ & 0.002& & & \\
   $\mathcal{Z}^{\mathcal{B}}_{\text{gu}}$ & 0.041 & &  & \\
   $\mathcal{Z}^{\mathcal{B}}_{\text{gg}}$ & 0.002 & & & \\
   $\mathcal{Z}^{\mathcal{B}}_{\text{gt}}$ & 0.001 & & & \\
   $\mathcal{Z}^{\mathcal{B}}_{\text{tu}}$ & 0.672 & & 13.9 & \\
   $\mathcal{Z}^{\mathcal{B}}_{\text{tg}}$ & 0.086 & & & \\
   $\mathcal{Z}^{\mathcal{B}}_{\text{tt}}$ & 0.034 & & & \\
  \hline
 \end{tabular}
 \caption{As Table \ref{tab:evidenceBF} but for Merging Galactic DNS systems. In this case, we find $\mathcal{Z^{B}}/\mathcal{Z^{A}}=3.3$.}
  \label{tab:evidenceBFM}
\end{table}

\begin{table}[!htb]
\centering
 \begin{tabular}{l|c|c|c|c}
  \hline
 Subhypothesis &  & $\mathcal{Z}^{\mathcal{A}}_{\text{u}}$ & $\mathcal{Z}^{\mathcal{A}}_{\text{g}}$ & $\mathcal{Z}^{\mathcal{A}}_{\text{t}}$ \\
  \hline
  & RE & 0.705 & 0.204 & 0.091 \\
  \hline
  $\mathcal{Z}^{\mathcal{B}}_{\text{uu}}$ & 0.667 & 4.8 &  & \\
   $\mathcal{Z}^{\mathcal{B}}_{\text{ug}}$ & 0.089 & & & \\
   $\mathcal{Z}^{\mathcal{B}}_{\text{ut}}$ & 0.041& & & \\
   $\mathcal{Z}^{\mathcal{B}}_{\text{gu}}$ & 0.092 & &  & \\
   $\mathcal{Z}^{\mathcal{B}}_{\text{gg}}$ & 0.013 & & & \\
   $\mathcal{Z}^{\mathcal{B}}_{\text{gt}}$ & 0.007 & & & \\
   $\mathcal{Z}^{\mathcal{B}}_{\text{tu}}$ & 0.075 & & & \\
   $\mathcal{Z}^{\mathcal{B}}_{\text{tg}}$ & 0.011 & & & \\
   $\mathcal{Z}^{\mathcal{B}}_{\text{tt}}$ & 0.005 & & & \\
  \hline
 \end{tabular}
 \caption{As Table \ref{tab:evidenceBF} but for Nonmerging Galactic DNS systems. In this case, we find $\mathcal{Z^{B}}/\mathcal{Z^{A}}=1.7$.}
  \label{tab:evidenceBFNM}
\end{table}

\subsection{Model Selection}
We organize our results three ways.
In Table~\ref{tab:evidenceBF}, we present the BFs for all 17 Galactic DNS systems (merging and nonmerging).
In Tables \ref{tab:evidenceBFM} and \ref{tab:evidenceBFNM}, we present the BFs for the 10 merging and 7 nonmerging systems, respectively. Here the merging (nonmerging) category includes those systems that will (not) merge within a Hubble time (see Table \ref{tab:BNSmass}).
We present hyperposteriors in the Appendix for the best subhypothesis found in Table~\ref{tab:evidenceBF}.
We highlight several take-away messages:
\begin{itemize}
    \item The data exhibits positive support for the hypothesis that the recycled NS mass and the slow NS mass are distributed differently.
    Using the entire sample, we find $\text{BF}=9$ for hypothesis $\mathcal{B}$ over $\mathcal{A}$.
    The best subhypothesis ($\mathcal{Z}^{\mathcal{B}}_{\text{tu}}$) is a two-Gaussian distribution for $m_r$ and a uniform distribution for $m_s$, with an RE of 64\%.
    We find it is strongly favored (BF=29) over the subhypothesis $\mathcal{Z}^{\mathcal{A}}_{\text{g}}$ (the conventional-wisdom model), which states that $m_r$ and $m_s$ are drawn from an identical Gaussian distribution.
    The second and third best subhypotheses are $\mathcal{Z}^{\mathcal{B}}_{\text{uu}}$ and $\mathcal{Z}^{\mathcal{B}}_{\text{gu}}$, having an RE of 20\% and 9\%, respectively.
    \item Some subhypotheses are strongly ruled out.
    For example, the combination of Gaussian $m_r$ and two-Gaussian $m_s$, or uniform $m_r$ and two-Gaussian $m_s$, yield an RE of $\sim 0.1\%$.
    \item The results do not change significantly if we analyze merging binaries as a special group.
    This is unsurprising because 9 out of 12 systems with precise component mass measurements are in this category.
    \item Data from nonmerging binaries is inconclusive.
    Presently, the BFs in the nonmerging table are too small to draw even preliminary conclusions.
    However, future observations could potentially highlight differences in the merging and nonmerging populations, for example, due to differences in the accretion process~\citep{Tauris17}.
\end{itemize}

\subsubsection{Discussion}
In regard to the difference in the distribution of $m_r$ and $m_s$, \citet{Ozel12} carried out a fit of the Gaussian distribution separately to the masses of pulsars and companions.
They found the most likely parameters are $\mu=1.35 M_{\sun}$, $\sigma=0.05 M_{\sun}$, and $\mu=1.32 M_{\sun}$, $\sigma=0.05 M_{\sun}$, for pulsars and companions, respectively. Therefore, they concluded that two members of DNS systems are drawn from the same underlying population. Although PSR J1906+0746, which is a slow pulsar, was mixed up with other recycled pulsars in \citet{Ozel12}, the difference between our result and theirs is due to the addition of new DNS systems with masses measured outside that ``typical" range. Some notable examples include the most massive $m_r$ at 1.64 $M_{\sun}$ (PSR J1913+1102), the lightest $m_s$ at $1.174 M_{\sun}$ (PSR J0453+1559), and the most massive $m_s$ at $1.46 M_{\sun}$ (PSR J0509+3801).
\citet{Tauris17} noticed a $\sim 0.1 M_{\sun}$ difference in the histogram of $m_r$ and $m_s$. 
They argued that this discrepancy is due to differences in the distribution of birth masses because recycled NSs in DNS systems are expected to gain $\leq 0.02 M_{\sun}$ from accretion.

The median values of marginalized posterior distributions under the best subhypothesis $\mathcal{Z}^{\mathcal{B}}_{\text{tu}}$ are $\{\mu_{1},\sigma_{1},\mu_{2},\sigma_{2},\alpha\}$ $=\{1.34 M_{\sun}, 0.02 M_{\sun}, 1.47 M_{\sun}, 0.15 M_{\sun}, 0.68\}$ for $m_r$, and $m_{s}^{l}=1.16 M_{\sun}$ and $m_{s}^{u}=1.42 M_{\sun}$.
We note that the secondary peak of the two-Gaussian distribution for $m_r$ is consistent with the distribution derived for recycled NSs with mostly white dwarf companions, $\mu=1.46 M_{\sun}$ and $\sigma= 0.21 M_{\sun}$, in \citet{Ozel12}.
If this bimodal structure is confirmed with the addition of more DNS measurements, it could lend support to the claim by \citet{Schwab10} that the NS mass spectrum has two peaks: one from Fe core collapse supernovae and one from electron capture supernovae\footnote{We caution that the mean values of our peaks appear to differ from those in~\citet{Schwab10}.}. Extending our analysis to all NS mass measurements may shed new light on this problem, as well as provide insights into the massive star evolution \citep{Raithel18}.

Recently, \citet{Huang18} analyzed the total mass measurements of 15 Galactic DNS systems, and found strong preference for a two-Gaussian distribution against the single-Gaussian model, using a simple likelihood ratio statistic.
\citet{Keitel18DNS} re-analyzed the same data and questioned the validity of such a claim. By performing various statistical tests, \citet{Keitel18DNS} suggested that there is no clear support for the two-Gaussian model and more data are required to draw firm conclusions.
Since we focus on the component mass measurements in this work, our results are not directly comparable to that of \citet{Keitel18DNS}. A common message from our work and \citet{Keitel18DNS} is that additional DNS discoveries are needed to settle the question of subpopulations.

\subsubsection{How Many Observations Are Required to Distinguish Two Populations?}

Here we determine how many observations are required to confidently distinguish the two distributions for $m_r$ and $m_s$, assuming they are indeed drawn from two different parent populations. We generate mock mass measurements of DNS systems assuming that $m_r$ follows a two-Gaussian distribution, and $m_s$ follows a uniform distribution, with parameters taking the median values of marginalized posterior distributions.

The mass measurement uncertainties of DNS systems to be discovered in future pulsar surveys are difficult to predict. Uncertainties of existing measurements vary by orders of magnitude, as can be seen in Figure \ref{fig:m12BNS}. For simplicity and conservativeness, we assume a measurement error of 0.04 $M_{\sun}$ for both $m_r$ and $m_s$, and a smaller error of 0.014 $M_{\sun}$ for $M_T$, similar to those of PSR J1913+1102, for all future discoveries. We further assume the posterior probability $p(m_r, m_s | d)$ peaks at injected values of $m_r$ and $m_s$ for all the mock measurements. In reality, measurements do not always produce posteriors that peak at the true parameter values. This would make our calculations somewhat optimistic by a factor of $\lesssim 2$. Since the assumed mass measurement precision is conservative compared to those in Table \ref{tab:BNSmass}, we expect our results to be realistic when interpreted as the required number of DNS systems with at least two post-Keplerian parameters measured. What is needed is the discovery of new DNS systems and measurement of their component masses through follow-up radio timing observations.

\begin{figure}[ht]
\begin{center}
  \includegraphics[width=0.46\textwidth]{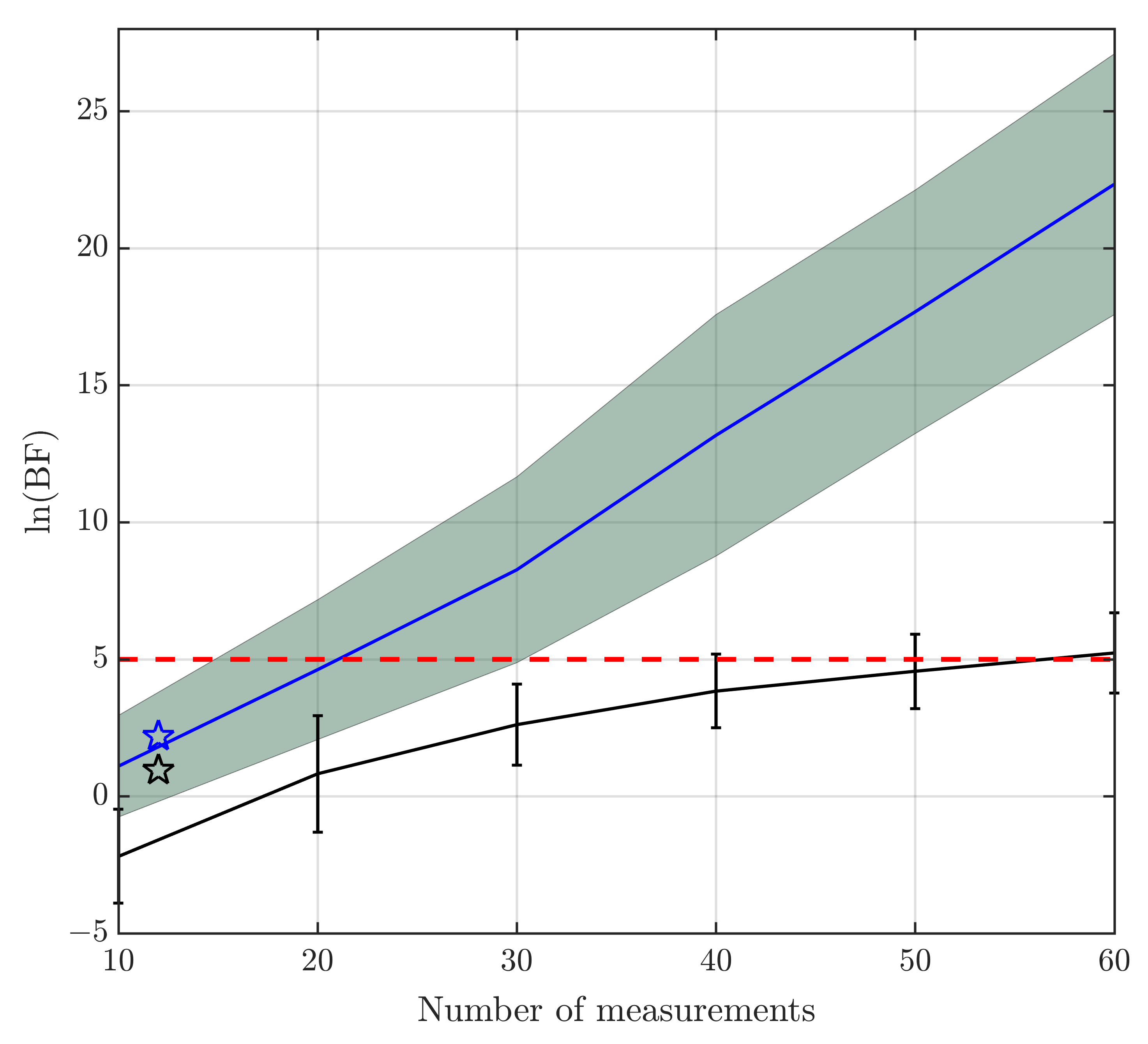}
  \caption{The natural logarithm of Bayes factor, ln(BF), between the hypothesis $\mathcal{B}$ and $\mathcal{A}$ (blue), and between the true subhypothesis and the next best subhypothesis within hypothesis $\mathcal{B}$ (black), as a function of the number of measurements. Solid curves and error regions/bars show median predictions and $1-\sigma$ confidence intervals, respectively, from simulated observations. The stars are computed from existing observations. \label{fig:lnBF}}
\end{center}
\end{figure}

Figure \ref{fig:lnBF} shows the natural logarithm of BF as a function of the number of DNS discoveries.
The solid blue curve with shaded region shows the BF between the hypotheses $\mathcal{B}$ and $\mathcal{A}$. The solid black curve with error bars shows the BF between the true subhypothesis (two-Gaussian $m_r$, and uniform $m_s$), and the second best subhypothesis, or the best subhypothesis when the true subhypothesis does not produce the highest evidence\footnote{This is usually the case when there are fewer than 20 measurements.}, within hypothesis $\mathcal{B}$. The red horizontal line marks the threshold of $\ln{(\text{BF})}=5$ as required for confident model selection. One can see that, on average, $\approx 20$ measurements are required to tell that $m_r$ and $m_s$ follow different distributions, and about 60 measurements are required to confidently distinguish the true subhypothesis from the other eight subhypotheses. The blue and black stars show results from existing observations, which are in good agreement with the simulations.

\begin{figure}[ht]
\begin{center}
  \includegraphics[width=0.46\textwidth]{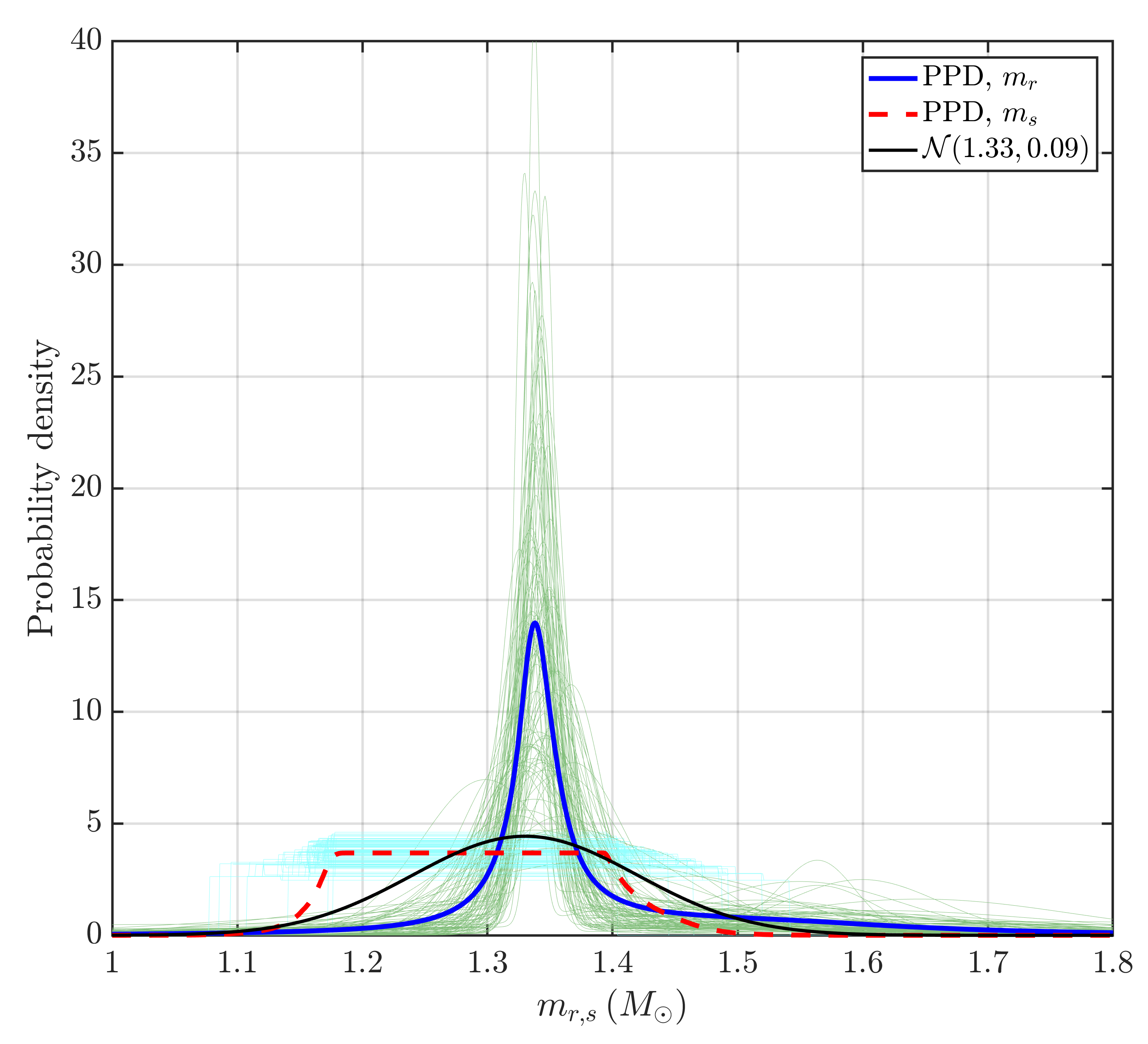}
  \caption{Posterior predictive distributions (PPD) of recycled NS mass $m_r$ and slow NS mass $m_s$. Thin curves are the distributions for 100 independent posterior samples. In comparison is the conventional Gaussian distribution with a mean of 1.33 $M_{\sun}$ and a width of 0.09 $M_{\sun}$. \label{fig:PPD_mrs}}
\end{center}
\end{figure}

\begin{figure}[!htb]
  \centering
  \includegraphics[width=0.46\textwidth]{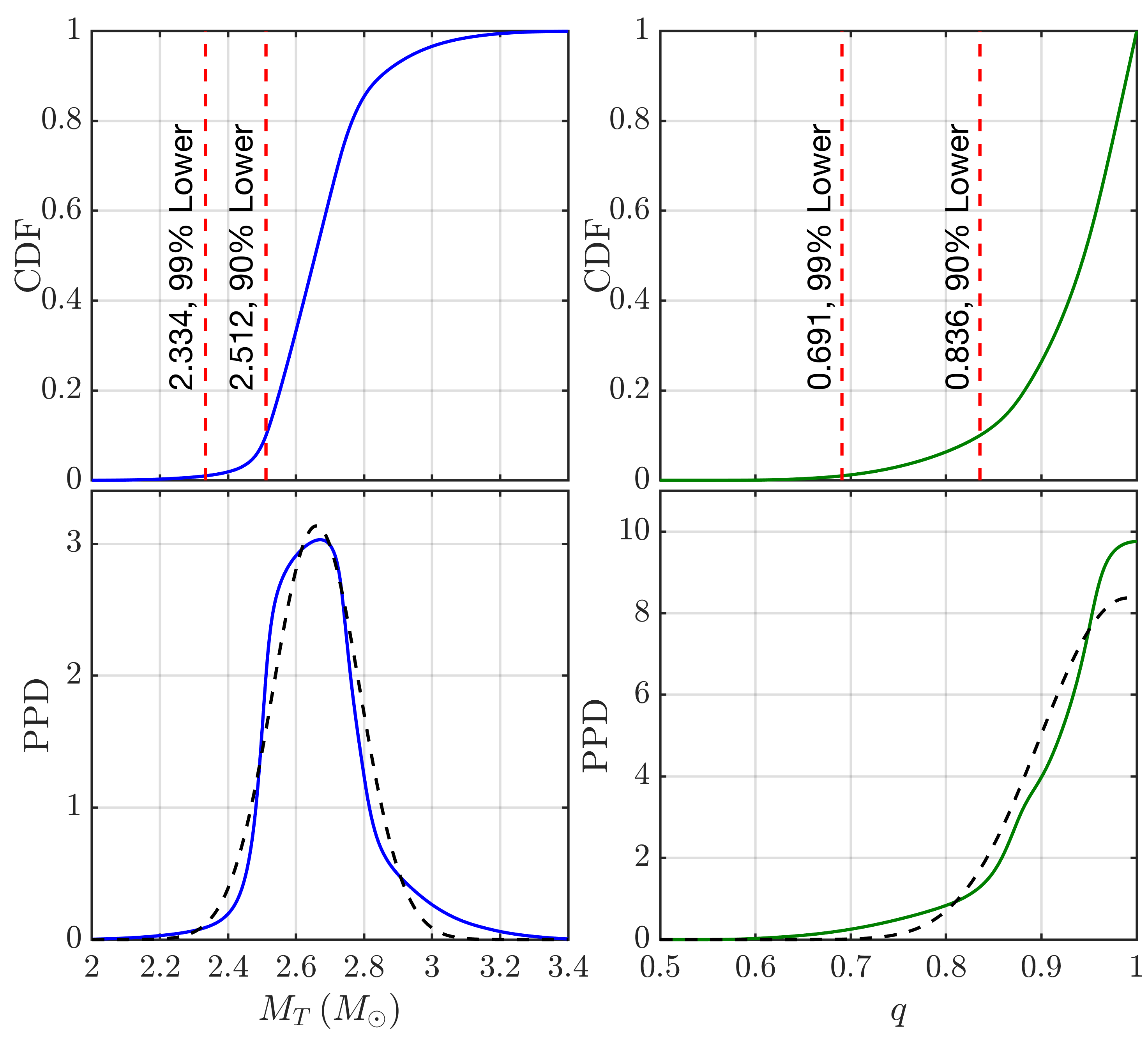}
  \caption{Posterior predictive distributions (PPD) and their cumulative distribution functions (CDF) for the binary total mass ($M_T$) and mass ratio ($q$). Solid curves are for the best subhypothesis found in this work, whereas dashed black curves in the lower panels show the distributions derived assuming both component NSs follow the conventional Gaussian distribution $\mathcal{N}(1.33,0.09)\, M_{\sun}$.}
  \label{fig:PPD_Mtq}
\end{figure}

\subsection{Posterior Predictive Distributions}

In this section, we present PPDs for the subhypothesis that best describes masses of all 17 Galactic DNS systems (two-Gaussian $m_r$ and uniform $m_s$). Figure \ref{fig:PPD_mrs} shows the PPDs for $m_r$ and $m_s$. For comparison, we also plot the conventional Gaussian distribution with a mean of 1.33 $M_{\sun}$ and a width of 0.09 $M_{\sun}$.

Distributions shown in Figure \ref{fig:PPD_mrs} can be converted to that for total mass and mass ratio, both of which are important for gravitational-wave data analysis. The distribution of total mass is also a key to telling how often do NS mergers produce black holes versus hypermassive NSs. Figure \ref{fig:PPD_Mtq} shows the PPDs (lower panels) and their cumulative distribution (upper panels) for $M_T$ and $q$. The red vertical lines in the upper panel mark the 99\% and 90\% confidence lower bounds. For example, we find that $M_T$ is above 2.334 $M_{\sun}$ and $q$ is larger than 0.691 with 99\% confidence. In contrast, these bounds become 2.364 $M_{\sun}$ and 0.779 for $M_T$ and $q$, respectively, if $m_r$ and $m_s$ are assumed to follow the conventional Gaussian distribution with $\mu=1.33 M_{\sun}$ and $\sigma=0.09 M_{\sun}$.

\section{Conclusions}
\label{sec:conclu}
The number of known DNS systems in our Galaxy has nearly doubled in the past several years. Radio pulsar timing observations provide precise measurements of NS masses for many of these systems.
This growing sample allows tighter constraints to be placed on the mass distribution.
In this work, we make use of mass measurements of 17 Galactic DNS systems, including precise component mass measurements of 12 systems, and perform Bayesian inference and model selection on the mass distribution.
We find tentative evidence for two distinct mass distributions for recycled and slow NSs, and for the bimodality of the recycled NS mass distribution.
Furthermore, the conventional model which states that both component NSs follow an identical Gaussian distribution is disfavored with a BF of 29, compared to a two-Gaussian mass distribution for recycled NSs and uniform mass distribution for slow NSs.
If recycled NSs have a different mass distribution from slow NSs, this could hint at subtleties in our understanding of supernovae mechanisms and/or accretion processes.

We show that precise measurements of component masses of around 20 DNS systems are required to confidently tell---with ln(BF)$>5$---whether or not recycled and slow NSs follow different mass distributions.
To establish the detailed shape of the mass distribution, measurements of 60 binaries are needed. This might become possible in the next decade, through ongoing pulsar surveys and new surveys with the Five-hundred-meter Aperture Spherical Telescope \citep{FAST11} in China and MeerKAT \citep{MeerTime}, a precursor for the planned Square Kilometre Array\footnote{https://www.skatelescope.org/} in South Africa. The method developed here will prove useful for further review of the mass distribution as new DNS systems are discovered and their masses are measured.

In this work we have focused on radio pulsar observations.
Our analysis does not take into account selection effects, which means the distributions derived here do not necessarily match the mass distribution of DNSs at birth or merger.
The former can be predicted from population synthesis studies.
The latter is directly measurable in gravitational-wave observations and will soon be probed with the many new discoveries expected when Advanced LIGO and Advanced Virgo progress toward their design sensitivities in the next few years.
We plan to develop a framework that links the three populations, facilitating comparison of gravitational-wave mergers with the Galactic DNS population.

\acknowledgments
X.Z. and E.T. are supported by ARC CE170100004.
E.T. is additionally supported through ARC FT150100281.
We thank Matthew Bailes for helpful feedback about selection effects. We also thank the anonymous referee, David Keitel, and Harald Pfeiffer for useful comments on the manuscript.

\bibliography{ref}

\appendix

\begin{figure*}[!htb]
  \centering
  \includegraphics[width=\textwidth]{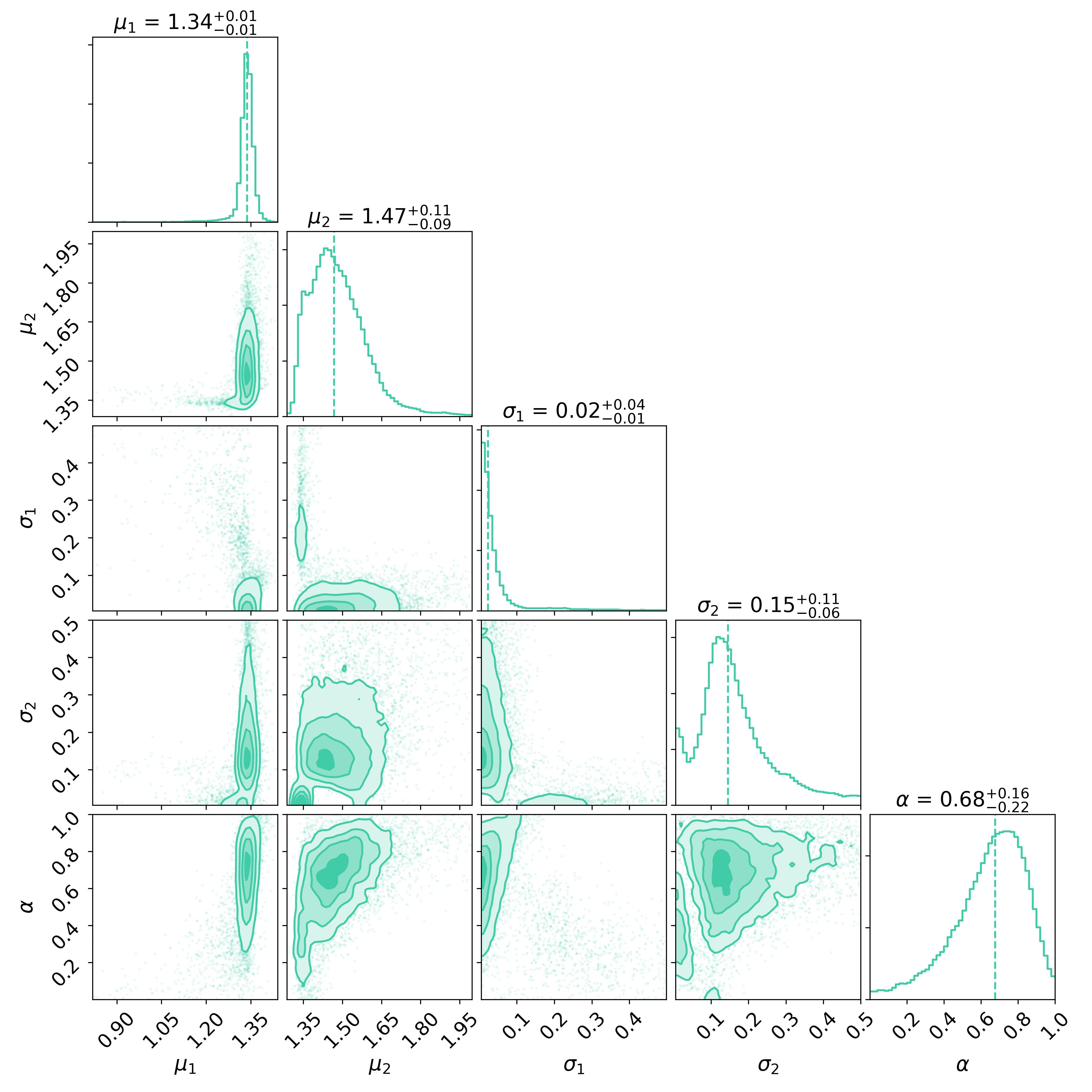}
  \caption{Posterior distributions of parameters for the two-Gaussian distribution for recycled NS mass $m_r$.}
  \label{fig:twoGcorner}
\end{figure*}

\begin{figure}[ht]
\begin{center}
  \includegraphics[width=0.48\textwidth]{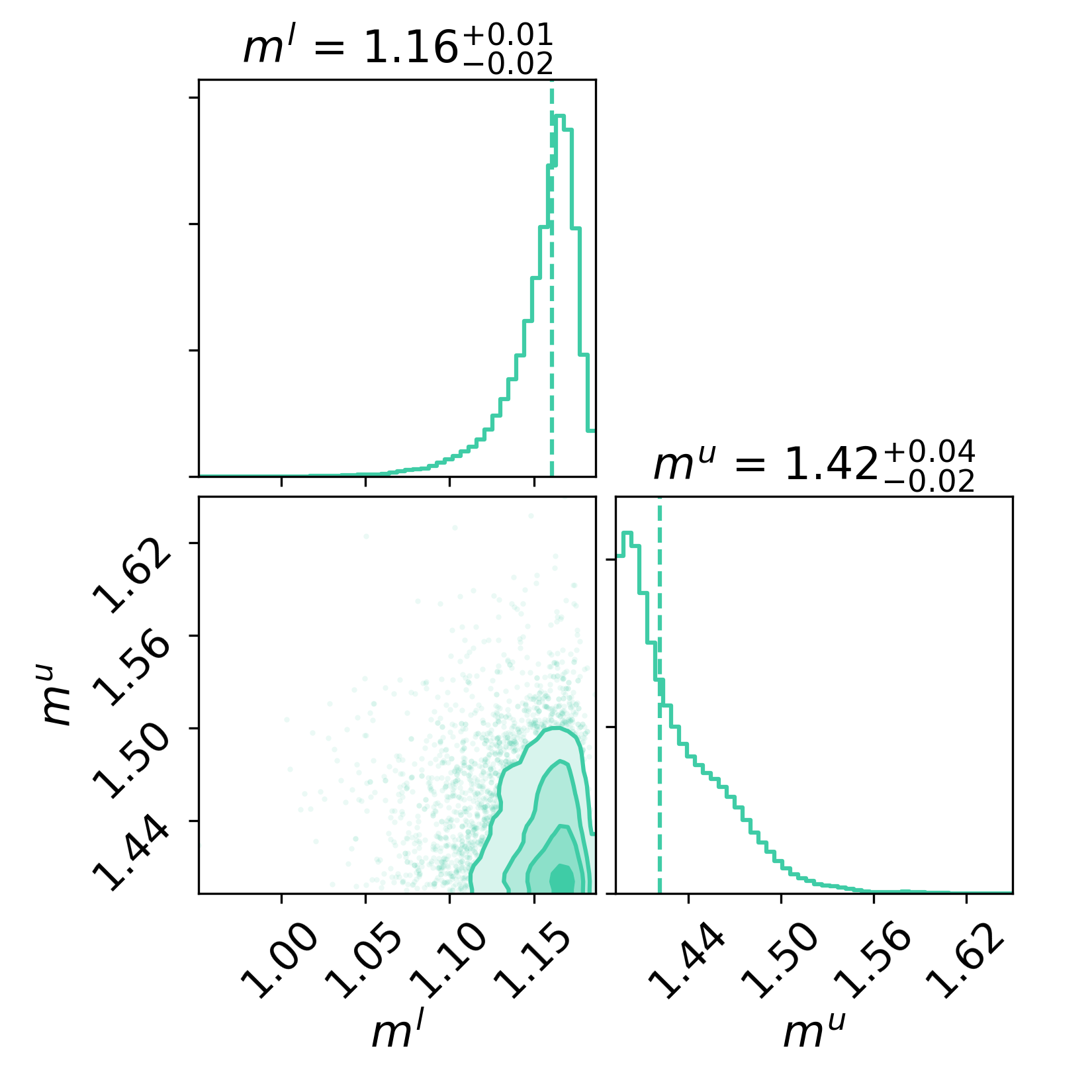}
  \caption{Posterior distributions of parameters for the uniform distribution for slow NS mass $m_s$.\label{fig:unifcorner}}
\end{center}
\end{figure}

\end{document}